\pdfoutput=1
\documentclass[11pt,twoside,a4paper,cmspaper,final,collab]{cms-tdr}
\def\svnVersion{582cd9a}\def\svnDate{2024/12/05}\def\cmsCernNoTag{CERN-EP-2024-186}\def\cmsCernDate{\today}\def\cmsMessage{Published in Physics Letters B as \href{http://dx.doi.org/10.1016/j.physletb.2024.139173}{\doi{10.1016/j.physletb.2024.139173}.}}
\begin{document}\cmsNoteHeader{HIG-23-003}

\newlength\cmsTabSkip\setlength{\cmsTabSkip}{1ex}
\newcommand{\VH}{\ensuremath{\PV\PH}\xspace}
\newcommand{\ttH}{\ensuremath{\ttbar\PH}\xspace}
\newcommand{\bbarH}{\ensuremath{\bbar\PH}\xspace}
\newcommand{\bbH}{\ensuremath{\bbar\PH}\xspace}
\newcommand{\bH}{\ensuremath{\PQb\PH}\xspace}
\newcommand{\bbHyt}{\ensuremath{\bbarH(y_\PQt^2)}\xspace}
\newcommand{\bbHyb}{\ensuremath{\bbarH(y_\PQb^2)}\xspace}
\newcommand{\bbHybyt}{\ensuremath{\bbarH(y_\PQt y_\PQb)}\xspace}
\newcommand{\ggH}{\ensuremath{\Pg\Pg\PH}\xspace}
\newcommand{\tH}{\ensuremath{\PQt\PH}\xspace}
\newcommand{\mH}{\ensuremath{m_{\PH}}\xspace}
\newcommand{\mt}{\ensuremath{m_{\mathrm{T}}}\xspace}
\newcommand{\Vjets}{\ensuremath{\PV\text{+jets}}\xspace}
\newcommand{\DYjets}{\ensuremath{\text{DY+jets}}\xspace}
\newcommand{\Wjets}{\ensuremath{\PW\text{+jets}}\xspace}
\newcommand{\mj}{\ensuremath{m_{\jH}}\xspace}
\newcommand{\mutau}{\ensuremath{\PGm\PGt_\mathrm{h}}\xspace}
\newcommand{\etau}{\ensuremath{\Pe\PGt_\mathrm{h}}\xspace}
\newcommand{\tautau}{\ensuremath{\PGt_\mathrm{h}\PGt_\mathrm{h}}\xspace}
\newcommand{\emu}{\ensuremath{\Pe\PGm}\xspace}
\newcommand{\HTT}{\ensuremath{\PH\to\PGt\PGt}\xspace}
\newcommand{\HWW}{\ensuremath{\PH\to \PW\PW}\xspace}
\newcommand{\bbar}{\ensuremath{\PQb\PAQb}\xspace}
\newcommand{\tauell}{\ensuremath{\PGt_{\ell}}\xspace}
\newcommand{\mtautau}{\ensuremath{m_{\PGt\PGt}}\xspace}
\newcommand{\mttot}{\ensuremath{m_{\mathrm{T,tot}}}\xspace}
\newcommand{\ptt}{\ensuremath{p_{\mathrm{T},\PGt\PGt}}\xspace}
\newcommand{\mcol}{\ensuremath{m_{\text{coll}}}\xspace}
\newcommand{\mvis}{\ensuremath{m_{\text{vis}}}\xspace}
\newcommand{\kapt}{\ensuremath{\kappa_\PQt}\xspace}
\newcommand{\kapb}{\ensuremath{\kappa_\PQb}\xspace}
\newcolumntype{L}[1]{>{\raggedright\let\newline\\\arraybackslash\hspace{0pt}}m{#1}}
\newcolumntype{C}[1]{>{\centering\let\newline\\\arraybackslash\hspace{0pt}}m{#1}}
\newcommand{\FF}{\ensuremath{F_{\mathrm{F}}}\xspace}
\newcommand{\jtotauh}{\ensuremath{\mathrm{j}\to\tauh}\xspace}
\newcommand{\pzeta}{\ensuremath{p_{\zeta}}\xspace}
\renewcommand{\ptvecmiss}{\ensuremath{\ptvec^{\text{ miss}}}\xspace}
\newcommand{\pttotvec}{\ensuremath{\ptvec^{\text{ tot}}}\xspace}
\newcommand{\pttot}{\ensuremath{p_{\mathrm{T,tot}}}\xspace}
\newcommand{\Dzeta}{\ensuremath{D_\zeta}\xspace}

\cmsNoteHeader{HIG-23-003}
\title{Search for bottom quark associated production of the standard model Higgs boson in final states with leptons in proton-proton collisions at \texorpdfstring{$\sqrt{s}=13\TeV$}{sqrt(s) = 13 TeV}}

\author*[desy]{Maryam Bayat Makou}
\author*[UZH]{Pascal Baertschi}
\author*[UZH]{Florencia Canelli}
\author*[desy]{Andrea Cardini}
\author*[UZH]{Adinda de Wit}
\author*[desy]{Oleg Filatov}
\author*[desy]{Elisabetta Gallo}
\author*[desy]{Rainer Mankel}
\author*[desy]{Alexei Raspereza}
\author*[haw]{Peer Stelldinger}
\author*[desy]{Roberval Walsh}

\date{\today}

\abstract{
  This Letter presents the first search for bottom quark associated production of the standard model Higgs boson, in final states with leptons. Higgs boson decays to pairs of tau leptons and pairs of leptonically decaying W bosons are considered. The search is performed using data collected from 2016 to 2018 by the CMS experiment in proton-proton collisions at a centre-of-mass energy of 13\TeV, corresponding to an integrated luminosity of 138\fbinv. Upper limits at the 95\% confidence level are placed on the signal strength for Higgs boson production in association with bottom quarks; the observed (expected) upper limit is 3.7 (6.1) times the standard model prediction.
}

\hypersetup{
pdfauthor={CMS Collaboration},
pdftitle={Search for the SM Higgs boson produced in association with bottom quarks in final states with leptons},
pdfsubject={CMS},
pdfkeywords={CMS, Higgs, tau, search, bottom}}

\maketitle

\section{Introduction}
\label{sec:1intro}
The discovery of the Higgs boson (H)~\cite{H125ATLAS,H125CMS,H125CMS2} by the ATLAS~\cite{ATLAS} and CMS~\cite{CMS:2008xjf} Collaborations in 2012 was a milestone in the study of the standard model (SM) of particle physics. A broad programme of measurements, studying the properties of this newly found particle, has been carried out over the subsequent decade~\cite{ATLAS:Hig102022,CMS:Hig102022}.

The production of the SM Higgs boson at the CERN LHC can occur via several mechanisms~\cite{LHCHiggsCrossSectionWorkingGroup:2016ypw}; the dominant ones, listed in order of decreasing production rate, are: gluon-gluon fusion (\ggH), vector boson fusion (VBF), Higgs-strahlung (\VH), top quark associated production (\ttH and \tH), and bottom quark associated production (\bbH and \bH, jointly referred to as \bbH in this Letter). 
With the exception of the \bbH and \tH processes, these production modes have all been observed~\cite{ATLAS:Hig102022,CMS:Hig102022}.
With a predicted production cross section of $0.48_{-0.11}^{+0.10}\unit{pb}$, calculated assuming a Higgs boson mass of 125.38\GeV in the four-flavour-scheme (4FS) and 5FS with Santander matching~\cite{LHCHiggsCrossSectionWorkingGroup:2016ypw,Dawson_2004,Dittmaier_2004,Harlander_2003,Forte_2015,Forte_2016,Bonvini:2015pxa,Bonvini:2016fgf}, \bbH production occurs almost as frequently as \ttH production. However, distinguishing the Higgs boson signal from backgrounds is experimentally more challenging for \bbH production than for \ttH production, since the top quark decay products provide a clear experimental signature. As such, \ttH production was already observed by both ATLAS~\cite{ATLAS:tth2018} and CMS~\cite{CMS:tth2018} in 2018. 
In addition to challenges related to the suppression of backgrounds, the \bbH process interferes with other Higgs boson production mechanisms~\cite{Pagani:2020rsg}. The destructive interference with Higgs boson production via \ggH, in particular, hinders the possibility of directly constraining the Yukawa coupling between the Higgs boson and bottom quarks through studies of this process. Measurements of Higgs boson decays to bottom quarks provide more precise constraints on this Yukawa coupling, while \bbH production remains sensitive to the correlation between the Higgs boson couplings to bottom and top quarks.

This Letter presents constraints on the production cross section of the Higgs boson in association with \PQb quarks, using the proton-proton ($\Pp\Pp$) collision data set recorded by the CMS detector at $\sqrt{s}=13\TeV$ in 2016--2018.
The search primarily targets Higgs boson decays into pairs of tau leptons. Higgs boson decays to leptonically decaying $\PW$ bosons are also considered in the electron-muon pair final state. These decay channels are chosen for their sizeable branching fractions, in combination with moderate backgrounds, which result from the presence of leptons in the final state. Compared with previous searches performed by the CMS experiment~\cite{CMS:2018rmh,CMS:2022goy}, this search is optimized to target the production mechanism of the 125\GeV Higgs boson in the SM.
Tabulated results are provided in the HEPData record for this analysis~\cite{hepdata}.

\begin{figure*}[htb]
    \centering
    \includegraphics[width=0.8\textwidth]{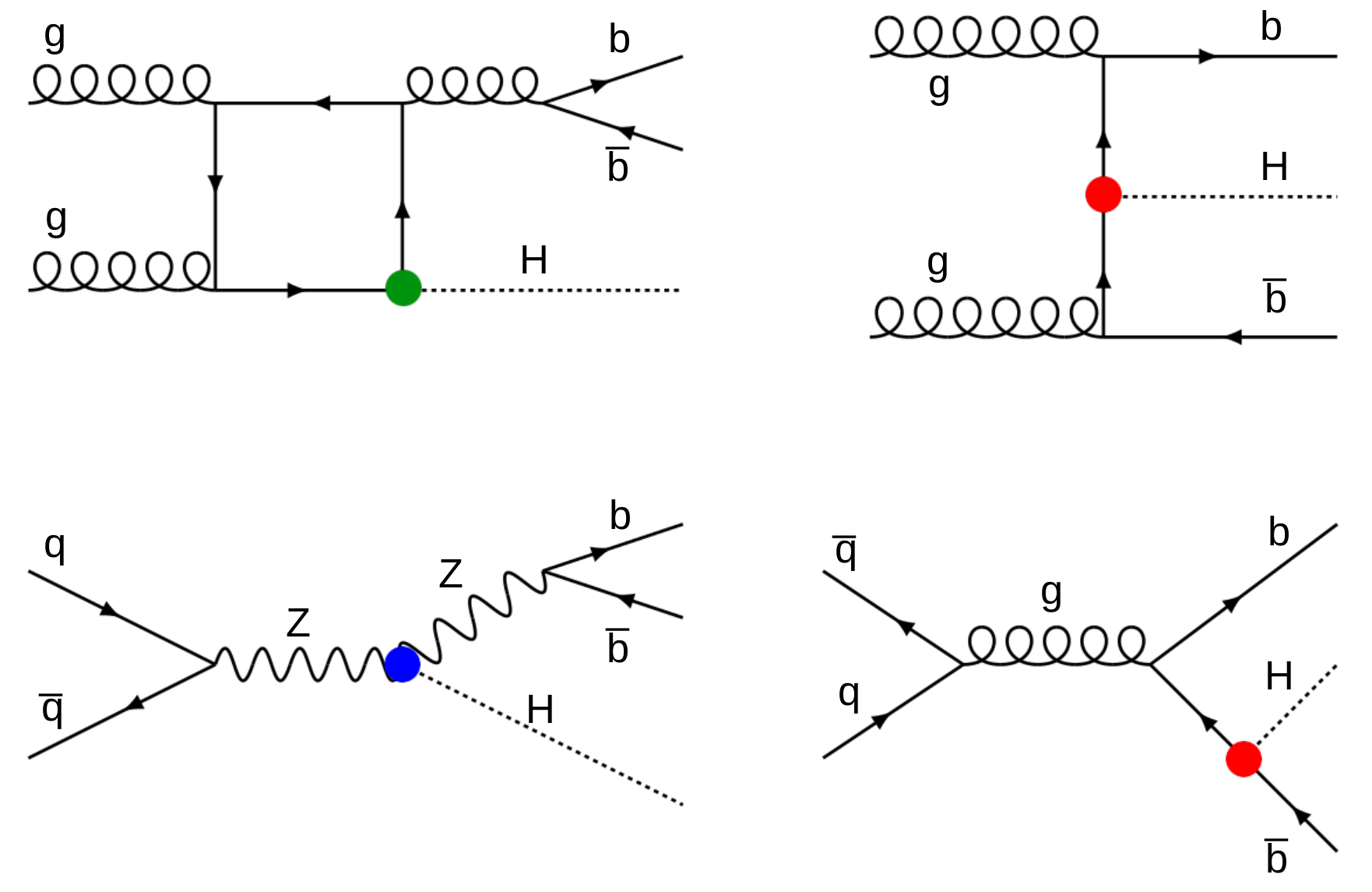}
    \caption{Dominant Feynman diagrams contributing to Higgs boson production in association with \PQb quarks in the four-flavor-scheme~\cite{Pagani:2020rsg,Deutschmann:2018avk}. The diagrams initiated by gluons (quarks) are shown in the upper (lower) row. The red circle is used to mark the Higgs boson coupling to \PQb quarks, the green circle marks the Higgs boson coupling to top quarks, and the blue circle marks the coupling between the Higgs boson and vector bosons. In the \ggH diagram (upper left), the additional gluon is radiated from within the quark loop, although it can equivalently radiate from one of the initial-state gluons.}
    \label{fig:feynman}
\end{figure*}

\section{Higgs boson production mechanisms and analysis strategy}
\label{sec:2bbH}

Figure~\ref{fig:feynman} shows the main Feynman diagrams for Higgs boson production in association with \PQb quarks. The upper left diagram represents \ggH production with an additional gluon splitting into a \bbar pair. In this diagram, the Higgs boson is produced via a quark (\PQq) loop that mainly involves the top quark. The production cross section is therefore dominated by a contribution proportional to the square of the top quark Yukawa coupling ($y_\PQt$). The theoretical production cross section for this process is evaluated as the fraction of the next-to-leading order (NLO) in quantum chromodynamics (QCD) \ggH production cross section, calculated in the 4FS~\cite{LHCHiggsCrossSectionWorkingGroup:2016ypw}, in which the gluon converts into a \bbar pair. This diagram interferes destructively with the diagram in Fig.~\ref{fig:feynman} (upper right), the production of a Higgs boson and a \bbar pair via bottom quark fusion. The production cross section of this process is proportional to the square of the bottom quark Yukawa coupling ($y_\PQb$).
Quark-initiated diagrams are shown in Fig.~\ref{fig:feynman} (lower): the Higgs boson is produced via the trilinear coupling to vector bosons (\PH\PV\PV) on the left and via the coupling $y_\PQb$ on the right. In the Higgs-strahlung diagram on the lower left, the \PQb quarks are produced through the decay of an on-shell \PZ boson, and the diagram includes only electroweak terms at leading order (LO). The resulting interference with the diagram on the lower right is therefore negligible with respect to the production cross section of the individual processes. In this Letter, the Higgs-strahlung process ($\Pp\Pp\to \PZ(\to \PQb\PAQb)+\PH$) is treated as a background because of the different coupling structure. The same treatment is reserved for the VBF production process with one or two outgoing $\PQb$ quarks.

To distinguish the different contributions to Higgs boson production in association with \PQb quarks, we use the following notation:
\begin{itemize}
    \item \bbHyb refers to \PQb quark associated Higgs boson production via Yukawa coupling to bottom quarks (e.g. the Feynman diagrams in Fig.~\ref{fig:feynman} (right)); 
    \item \bbHyt marks Higgs boson production via a top quark loop; and
    \item \bbHybyt labels the interference term between the top-quark-mediated and the bottom-quark-mediated production processes.
\end{itemize}
The inclusive production across these terms is labelled as \bbH or $\bbH(y_\PQb,y_\PQt)$.

The SM prediction for the \bbH production cross section is therefore split into three contributions with the following production cross sections~\cite{LHCHiggsCrossSectionWorkingGroup:2016ypw,Dawson_2004,Dittmaier_2004,Harlander_2003,Forte_2015,Forte_2016,Bonvini:2015pxa,Bonvini:2016fgf,Pagani:2020rsg,Deutschmann:2018avk}: 1.040\unit{pb} for \bbHyt (NLO reweighted to next-to-next-to-NLO (N$^3$LO) in QCD and NLO in electroweak), 0.482\unit{pb} for \bbHyb (NLO in QCD), and $-0.033\unit{pb}$ for \bbHybyt (NLO in QCD), amounting to a total predicted production cross section of $\sigma(y_\PQb,y_\PQt)=1.489\unit{pb}$.

The \bbH search presented in this Letter is performed in final states of two oppositely charged leptons with different flavour, or hadronically decaying tau lepton (\tauh) candidates:
\begin{itemize}
    \item two hadronically decaying tau leptons (\tautau);
    \item an electron and a hadronically decaying tau lepton (\etau);
    \item a muon and a hadronically decaying tau lepton (\mutau);
    \item an electron and a muon (\emu).
\end{itemize}
The first three final states originate from the decay of the Higgs boson to a pair of tau leptons, at least one of which decays hadronically. The \emu final state originates either from a Higgs boson decaying to a pair of tau leptons or from a Higgs boson decaying to a pair of \PW bosons. These two decay chains result in a similar event topology, and are therefore investigated as a single analysis channel. Final states with two electrons or muons have not been considered due to the overwhelming Drell-Yan background and low branching fractions.

\section{The CMS detector}
\label{sec:3CMS}
The CMS apparatus~\cite{CMS:2008xjf} is a multipurpose, nearly hermetic detector, designed to trigger on~\cite{CMS:2020cmk,CMS:2016ngn} and identify electrons, muons, photons, and hadrons~\cite{CMS:2020uim,CMS:2018rym,CMS:2014pgm} produced in proton-proton collisions. A ``particle-flow'' (PF) algorithm \cite{CMS:2017yfk} aims to reconstruct all individual particles in an event, combining information provided by the all-silicon inner tracker and by the crystal electromagnetic (ECAL) and brass-scintillator hadron calorimeters, operating inside a 3.8\unit{T} superconducting solenoid, with data from the gas-ionization muon detectors embedded in the flux-return yoke outside the solenoid. The reconstructed particles are used to build tau leptons, jets, and missing transverse momentum~\cite{CMS:2018jrd,CMS:2016lmd,CMS:2019ctu}.
The missing transverse momentum vector \ptvecmiss is computed as the negative vector sum of the transverse momenta of all the PF candidates in an event, and its magnitude is denoted as \ptmiss~\cite{CMS:2019ctu}. The \ptvecmiss is modified to account for corrections to the energy scale of the reconstructed jets in the event.

Events of interest are selected using a two-tiered trigger system. The first level, composed of custom hardware processors, uses information from the calorimeters and muon detectors to select events at a rate of around 100\unit{kHz} within a fixed latency of about 4\mus~\cite{CMS:2020cmk}. The second level, known as the high-level trigger (HLT), consists of a farm of processors running a version of the full event reconstruction software optimized for fast processing and reduces the event rate to around 1\unit{kHz} before data storage~\cite{CMS:2016ngn}.

A more detailed description of the CMS detector, together with a definition of the coordinate system used and the relevant kinematic variables, can be found in Ref. \cite{CMS:2008xjf}.

\section{Data and simulated samples}
\label{sec:4Samples}
The $\Pp\Pp$ collision data set used in this search was recorded by the CMS experiment between 2016 and 2018, at a centre-of-mass energy of 13\TeV. It corresponds to a total integrated luminosity of 138\fbinv.

Several different Monte Carlo (MC) event generators are used to produce samples of simulated events of the signal and background processes.
The simulation of the \bbH signal was performed at NLO in QCD using the \MGvATNLO 2.6.1 generator~\cite{bib:MADGRAPH}.
Three samples of simulated events are used, corresponding to the three distinct Higgs boson couplings contributing to the \bbH process ($y_\PQb^2$, $y_\PQt^2$, and $y_\PQt y_\PQb$). A sample of \ggH events is simulated at NLO in QCD using the \MGvATNLO 2.6.5 generator. Other Higgs boson production mechanisms, including VBF, \VH, and \ttH are generated at NLO in QCD with the \POWHEG 2.0 generator~\cite{powheg,Alioli:2010xd,Nason:2004rx}.

In all channels, the SM background is dominated by top quark-antiquark pair production (\ttbar) and by the production of a leptonically decaying weak boson in association with jets (\Vjets). A sample of simulated \ttbar events is generated with the \POWHEG 2.0 generator at NLO in QCD. The production of \Vjets events is simulated with the \MGvATNLO 2.6.5 generator at NLO in QCD. Samples of single top quark production events are generated at NLO in QCD with the \POWHEG 2.0 generator, while diboson events are produced at NLO in QCD with the \MGvATNLO 2.6.5 generator, or with the \POWHEG generator, depending on the specific combination of vector bosons studied. 
The background from SM events composed solely of jets produced through the strong interaction, referred to as QCD multijet events, is estimated from data as described in Section~\ref{sec:6bkgandSF}.

The NNPDF 3.1~\cite{Ball_2017} set of parton distribution functions (PDFs) at next-to-NLO (NNLO) in QCD is used in all simulated samples. Parton showering and hadronization are performed with \PYTHIA 8.240 with the CP5~\cite{pythia8,pythia8_tuneCP5} underlying event tune. The CMS detector response simulation is performed with \GEANTfour~\cite{bib:GEANT4} for all processes.
In the samples produced at NLO in QCD with \MGvATNLO, the FxFx jet merging scheme is employed~\cite{Frederix:2012ps}. 

Event reconstruction in simulated samples and recorded data is performed with the same software. The recorded data samples contain additional $\Pp\Pp$ interaction vertices from the same or nearby bunch crossings (pileup). Such additional interaction vertices are generated with \PYTHIA and added to all simulated events in accordance with the expected pileup distribution.
Corrections are applied to the simulated samples to match the distribution of the pileup multiplicity measured in the recorded data for each year.

\section{Trigger and event selection}
\label{sec:5EvSel}

The online selection~\cite{CMS:2016ngn} for the \tautau channel requires two hadronically decaying tau leptons. In the \mutau (\etau) channel the presence of a muon (electron) is required, and in the \emu channel a muon-electron pair must be present. In all channels, the selected leptons ($\ell$) or hadronically decaying tau leptons (\tauh) are required to match the objects used in the online event selection.

The following selection criteria are common across all channels:
\begin{itemize}
    \item there must be a \tautau, \etau, \mutau, or \emu pair with opposite electric charge. No additional electrons or muons, passing loose identification criteria, may be present in the event;
    \item the leptons and \tauh candidates must be separated by $\Delta R=\sqrt{\smash[b]{(\Delta\phi)^2+(\Delta\eta)^2}}> 0.5 (0.3)$ in the \tautau, \etau, and \mutau (\emu) channels. Here, $\phi$ and $\eta$ refer to the azimuthal angle and the pseudorapidity of the object, respectively;
    \item there must be either one or two jets identified as originating from a \PQb quark.
\end{itemize}

Across all channels, jets are selected if they have $\pt > 20\GeV$, $\abs{\eta} < 2.4$, and if they pass quality criteria based on the jet shape. Jet constituents coming from pileup interactions are subtracted.
Jets originating from \PQb quarks are identified with the \textsc{DeepJet} algorithm~\cite{DeepJet}, which uses information on tracks and secondary vertices associated with the jet to identify its type (\Pg, \PQu/\PQd/\PQs, \PQc, or \PQb).
In this Letter an efficiency of 75\% for \PQb tagging is chosen, corresponding to a mistagging probability for jets originating from gluons or light quarks of about 3\%.

In the \tautau channel, two isolated \tauh candidates with $\pt>40 \GeV$ and $\abs{\eta}<2.1$ are required.
To distinguish genuine \tauh decays from jets originating from the hadronization of quarks or gluons, and from electrons or muons, the \textsc{DeepTau} algorithm~\cite{CMS:2022prd} is used. Information from all individually reconstructed particles near the \tauh candidate axis is combined with properties of the \tauh candidate and of the event. Genuine \tauh candidates are selected with an efficiency of 70\% for a jet misidentification rate of approximately 0.43\%, averaging across jet \pt and flavour. The misidentification rates of electrons and muons as \tauh candidates are approximately 3\% and 0.2\%.

In the \etau, \mutau, and \emu channels, electrons and muons coming from \PW boson or tau lepton decays are treated according to the standard CMS reconstruction for electrons and muons~\cite{CMS:2020uim,CMS:2018rym}.

In the \etau and \mutau channels, the leptonically decaying tau lepton (\tauell) must have \pt at least 1\GeV above the trigger threshold. The electron is required to have $\abs{\eta}<2.1$ and $\pt >26\,(33) \GeV$ in the analysis of 2016 (2017, 2018) data, while the muon is required to have $\abs{\eta}<2.1$ and $\pt >23\,(25) \GeV$ for the 2016 (2017, 2018) data-taking period. Quality criteria are used to select the leptons based on their relative isolation from surrounding hadronic activity and the reconstruction quality of their tracks and energy deposits in the CMS subdetectors.
The \tauh candidates are required to have $\pt > 30\GeV$, $\abs{\eta} < 2.3$, and to satisfy the same quality criteria as are applied in the \tautau channel. To suppress the Drell--Yan background contribution, we use the \textsc{DeepTau} algorithm to reject electrons and muons that mimic \tauh signatures. The residual electron (muon) misidentification rate is 3\% ($<0.1\%$)~\cite{CMS:2022prd}.
The invariant mass between the \ptvecmiss and the lepton \ptvec, calculated as
\begin{equation}
\label{eq:mT}
\mt(\ptvec,\ptvecmiss) = \sqrt{{2\pt\ptmiss(1-\cos[\Delta\phi(\ptvec,\ptvecmiss)])}},
\end{equation}
is required to be smaller than 60\GeV to reduce the contribution from \Wjets events. This also results in the rejection of \HWW events; the residual contribution of such events is around 50 times smaller than the \HTT signal yield in the \etau and \mutau channels.

In the \emu channel, both the electron and the muon are required to have $\pt>15\GeV$ and $\abs{\eta}<2.4$, and to pass loose reconstruction criteria.
As a result of the kinematical similarities for electron-muon pairs originating from \HWW and \HTT decays, as well as the overlap in acceptance, these decays were studied in a single analysis channel.

\section{Background estimation}
\label{sec:6bkgandSF}

A multitude of background processes contribute to the presented search. Those backgrounds that include genuine hadronically decaying tau leptons, prompt leptons, or leptons misidentified as \tauh candidates are estimated from simulated events. Notable contributions to these backgrounds originate from \DYjets, diboson, and \ttbar production in the absence of jets misidentified as \tauh candidates. Corrections are derived in sideband regions to better describe the data recorded by the CMS experiment.
Processes involving jets misidentified as \tauh candidates and nonprompt leptons, appearing in multijet topologies, are estimated using control regions (CRs) in data, as discussed in the following section.

\subsection{Methods based on control regions in data}
In the \tautau, \etau, and \mutau final states, a large fraction of the background consists of jets misidentified as hadronic tau lepton decays. This contribution is estimated with the ``fake factor'' (\FF) method, as described in Ref.~\cite{CMS:HIG-15-017}. The shape of the misidentified \tauh background is estimated from a CR obtained by inverting the isolation criterion for the leading \tauh candidate. This CR is thus enriched in jets misidentified as \tauh candidates. The background shape is retrieved via event weights, denoted \FF, that are derived in determination regions (DRs) targeting a specific source of jet contamination. For the \tautau channel only a QCD multijet DR is used, whereas the \etau and \mutau channels require two additional DRs: one enriched in \Wjets events and one enriched in \ttbar events. The weights evaluated in these DRs are parameterized as a function of the transverse momenta of the \tauh candidates, and their angular separation, for the \tautau channel. In the \etau and \mutau channels, they are parameterized as a function of the \tauh candidate \pt. The parameterization in the \tautau channel also depends on other event properties, such as the number of gluon and light-flavour quark jets or \PQb jets, and the \ptmiss. The weighted sum of these \FF is then used to estimate the misidentified \tauh contribution to the signal region.

In the \emu channel, the QCD multijet background is estimated from events with a reconstructed electron-muon pair with the same electric charge. A transfer factor between this region, and the one where the leptons have opposite charge, is determined in events where the muon isolation requirement is inverted. The transfer factor is determined as a function of the lepton momenta, their angular separation, and the number of jets in the event. This background estimation method is commonly referred to as the ``ABCD'' method (discussed in Refs.~\cite{CDF:1995lcl,CMS:2017HTT}).

\subsection{Other backgrounds}
Other background processes are estimated from simulation. They are corrected to better describe the recorded data separately for each data-taking period.

Corrections are mostly determined in dedicated sideband regions, orthogonal to the event selection described in Section~\ref{sec:5EvSel}. For example, single-lepton trigger efficiencies are computed for the \etau and \mutau channels using $\PZ\to \ell\ell$ enriched CRs, while those for the combined triggers requiring two \tauh candidates or an electron and muon pair are determined, respectively, in a \mutau- and a \ttbar-enriched CR.
 
Simulation-to-data correction factors are applied to the simulated samples to correct for the mismodelling of the electron and muon identification and isolation efficiencies, and for the \tauh candidate identification efficiency and energy scale. The light lepton efficiencies are measured as a function of the lepton \pt and $\abs{\eta}$, using a ``tag-and-probe'' technique in $\PZ\to\Pe\Pe$ and $\PZ \to\PGm\PGm$ events~\cite{CMS:2018rym}. Correction factors for genuine \tauh candidates are determined based on their decay mode and \pt, while those for misidentified electrons or muons depend on $\eta$~\cite{CMS:2022prd}.

Jet energy and resolution corrections are applied to mitigate the effect of pileup interactions and detector inefficiencies during data taking~\cite{CMS:2020ebo}. Measurements of the momentum balance in dijet, $\PGg + \text{jet}$, $\PZ + \text{jet}$, and multijet events are used to correct for any residual differences in the jet energy scale between data and simulation~\cite{CMS:2016lmd}.

To each reconstructed jet with $\pt> 20 \GeV$, \PQb tagging scale factors are applied. The \PQb tagging score is weighted based on a multivariate classifier using several jet properties to improve the data modelling~\cite{CMS:2017wtu}.

The \PZ boson mass and momentum are corrected from $\PZ\to\PGm\PGm$ events as a function of the lepton \pt and $\eta$. The \pt spectrum in \ttbar events is weighted to better describe the data, following the method described in Ref.~\cite{CMS:2020mpn}.

\section{Event classification}
\label{sec:7BDT}
To improve the separation of signal from the background, boosted decision trees (BDTs) are used to classify events into distinct categories. Multiclass BDT models were trained for each channel and data-taking period. The output categories were chosen to reflect the different background compositions in each channel. 

\begin{table*}[!htb]
\centering
  \topcaption{Summary of the BDT categories defined for each channel.}
 \begin{tabular}{cC{2.7cm}C{2.2cm}C{2.6cm}}
 Channel & \emu & $\ell$\tauh & \tautau\\ \hline
 BDT Categories & DY,\; TT,\newline\vspace{1mm} \bbH($\to\PW\PW$), \newline\vspace{1mm} \bbH($\to\PGt\PGt$) & DY,\; TT,\newline\vspace{1mm} \bbH($\to\PGt\PGt$)  & DY+Higgs, TT,\newline\vspace{1mm} j$\to\tauh$ misid.,\newline\vspace{1mm} \bbH($\to\PGt\PGt$)\\  
 \end{tabular}
 \label{tab:BDTcat}
\end{table*}
\begin{table*}[!htb]
    \centering
    \topcaption{Input variables to the BDT classifiers used in each of the studied channels. Each variable is marked with the $\checkmark$ symbol if it is used for the training of the BDT models in a particular channel, or the $\times$ symbol if it is not used. Variables associated with the di-lepton system and serving as estimators for the Higgs boson properties are the most significant in the training.}
    \begin{tabular}{cccc}
        Variable & \emu & $\ell$\tauh & \tautau \\ \hline 
        $\mtautau$  & $\times$ & $\checkmark$ & $\checkmark$ \\
        $\ptt$  & $\times$ & $\checkmark$ & $\checkmark$ \\
        $\mttot$ & $\checkmark$ & $\times$ & $\times$\\ 
        $\pttot$ & $\checkmark$ & $\times$ & $\times$\\ 
        $\mvis$ & $\checkmark$ & $\checkmark$ & $\checkmark$ \\
        $\mcol$ & $\times$ & $\checkmark$ & $\times$ \\[\cmsTabSkip]
        $\mt$ & $\times$ & $\checkmark$ & $\times$ \\
        $\Dzeta$ & $\checkmark$ & $\checkmark$ & $\times$ \\
        \ptmiss & $\times$ & $\times$ & $\checkmark$ \\[\cmsTabSkip]
        Electron \pt & $\checkmark$ & $\times$ & $\times$ \\
        Muon \pt & $\checkmark$ & $\times$ & $\times$ \\
        \pt of leading \tauh & $\times$ & $\times$ & $\checkmark$ \\
        \pt of trailing \tauh & $\times$ & $\times$ & $\checkmark$ \\
        $\Delta  \eta$ between lepton and \tauh & $\times$ & $\checkmark$ & $\times$ \\[\cmsTabSkip]
        Number of \PQb-tagged jets & $\checkmark$ & $\times$ & $\checkmark$ \\
        \pt of leading \PQb-tagged jet & $\checkmark$ & $\checkmark$ & $\checkmark$ \\
        \pt of trailing \PQb-tagged jet  & $\times$ & $\checkmark$ & $\times$ \\
        B tag score for leading \PQb-tagged jet & $\times$ & $\checkmark$ & $\checkmark$ \\
        $\Delta\eta$ between di-$\PGt$ \pt and leading \PQb-tagged jet & $\times$ & $\checkmark$ & $\times$ \\
        B tag score for trailing \PQb-tagged jet & $\times$ & $\checkmark$ & $\checkmark$ \\[\cmsTabSkip]
        Number of jets & $\checkmark$ & $\times$ & $\checkmark$ \\
        \pt of leading jet & $\checkmark$ & $\times$ & $\checkmark$ \\
        \pt of trailing jet & $\checkmark$ & $\times$ & $\checkmark$ \\
        Dijet invariant mass & $\times$ & $\times$ & $\checkmark$ \\
        Dijet $\Delta\eta$ & $\checkmark$ & $\times$ & $\checkmark$ \\
        \end{tabular}
    \label{tab:BDTinput}
\end{table*}

For an event, each trained BDT model returns a number of output scores equal to the number of output categories. The scores are then normalized so that their sum is equal to 1. Each output score is therefore interpreted as the probability for the event to originate from one of the physical processes targeting the associated category. Events are sorted into the category for which the output score is the highest. 

All BDT models have a dedicated category targeting \HTT decays, trained on the \bbHyb and \bbHyt processes. The interference term was not used in the BDT training due to the low acceptance. Other common categories target the \DYjets (DY) and \ttbar (TT) background processes, as their signatures in the detector resemble those of \PQb quark associated Higgs boson production in final states involving tau leptons.
The \emu channel has an additional signal category, targeting Higgs boson decays to \PW bosons. As the sensitivity of this channel is driven by the \HWW process, the BDT is trained to optimize the separation between this signal and the \ttbar background.
In the \tautau channel the contribution of QCD multijet production with jets misidentified as \tauh candidates (j$\to\tauh$ misid.) to the total background is larger than in the other channels and was therefore assigned to a dedicated category.

Additional BDT models incorporating a further background category targeting other Higgs boson production mechanisms, such as the Higgs-strahlung process with a \PZ boson decaying into a pair of \PQb quarks, were trained. These models identified the \bbH process less efficiently, with a marginal improvement found only in the \tautau channel. After applying the event selection introduced in Section~\ref{sec:5EvSel}, only a small number of events is left in this Higgs category, which is therefore merged with the \DYjets category for the statistical inference.

A summary of the BDT categories used in each channel is shown in Table~\ref{tab:BDTcat}. Although the label ``TT'' is used in all categories, the \ttbar events used in each trained model differ: in the \tautau channel only events with genuine \tauh candidates are targeted by the ``TT'' category, in the other channels the ``TT'' categories target \ttbar events with no further selection.

As shown in Table~\ref{tab:BDTinput}, the set of variables used in the BDT training differs slightly across channels. Each set was chosen based on the potential separation between the \bbH signals from the various backgrounds in each channel. Variables that were not satisfactorily modelled in the MC samples used to train the BDT models were removed from a set to avoid introducing a bias in the relative channel towards a specific category. In cases where the mismodelling was specifically found for one year of data taking, the variable was nonetheless removed from the training set for all BDT models related to that channel.
Several input features are based on approximations of the Higgs boson invariant mass and transverse momentum. These event properties bring the best separation power between the \bbH process and the backgrounds. 
In the \etau, \mutau, and \tautau channels, a dedicated likelihood-based algorithm, \textsc{SVFit}~\cite{SVFit}, is used to reconstruct the kinematical properties of the di-tau system based on the momentum of the individual tau leptons, their decay modes, and the \ptmiss. 
The resulting di-tau mass and momentum are labelled as $\mtautau$ and $\ptt$. The kinematical properties of individual tau leptons are additionally used as separate input features in the \tautau channel.

Other variables related to the di-tau system are the collinear di-tau mass ($\mcol$), the invariant mass of the visible tau decay product system ($\mvis$), and the pseudorapidity separation between the visible decay products of the tau leptons ($\Delta  \eta_{\ell,\tauh}$).
The collinear mass is reconstructed assuming that the neutrinos produced in tau lepton decays are collinear with the visible tau lepton decay products. It is defined as~\cite{Elagin:2010aw}:
\begin{equation}
\mcol = \frac{\mvis}{\sqrt{{\frac{\pt^{\ell}}{\pt^{\ell}+\ptmiss}\frac{\pt^{\tauh}}{\pt^{\tauh}+\ptmiss}}}}.
\end{equation}
Both $\mcol$ and $\mtautau$ are insensitive to \HWW decays due to their different \ptmiss distribution.

A similar assumption for the neutrino direction of flight underpins the introduction of the $\Dzeta$ variable. This is defined as the linear combination of the projection of the \ptmiss along the bisector of the leptons or \tauh candidates ($\hat{\zeta}$) and the momentum of the dilepton (or lepton plus \tauh) system ($\pttotvec$):
\begin{equation}
\Dzeta=\pzeta^{\text{miss}}-0.85 \pzeta^\text{vis},
\end{equation} 
with 
\begin{equation}
\pzeta^\text{miss}=\ptvecmiss\cdot \hat{\zeta}\; \text{ and  }\; \pzeta^{\text{vis}}=\pttotvec\cdot \hat{\zeta}.
\end{equation} 
The value of 0.85 was optimized to maximize the separation power between resonant $\PGt_\Pe \PGt_{\PGm}$ decays from Higgs or \PZ bosons and the more isotropic neutrino emissions found in \ttbar or Higgs boson decays to \PW bosons~\cite{PhysRevLett.96.011802}.

The transverse mass (\mt), introduced in Eq.~(\ref{eq:mT}), is calculated using different objects, depending on the channel.
In the \etau and \mutau channels, the transverse mass is computed from the charged lepton and the missing transverse momentum. The observable is used to reduce the background contribution from electroweak processes in these channels. In the \emu channel, \mt is used to compute the total transverse mass of the electron, muon, and missing transverse momentum system:
\begin{equation}
\mttot = \sqrt{{\mt (\mu,\Pe)^2+\mt (\Pe,\ptmiss)^2+\mt (\mu,\ptmiss)^2}},
\end{equation}
with \pttot being the corresponding total transverse momentum. The two variables are used as estimators of the Higgs boson transverse mass and momentum for both the \HTT and \HWW decays. 

Other features that enter the BDT training relate to the reconstructed jets.

\begin{figure*}[!htb]
  \centering
    \includegraphics[width=.45\textwidth]{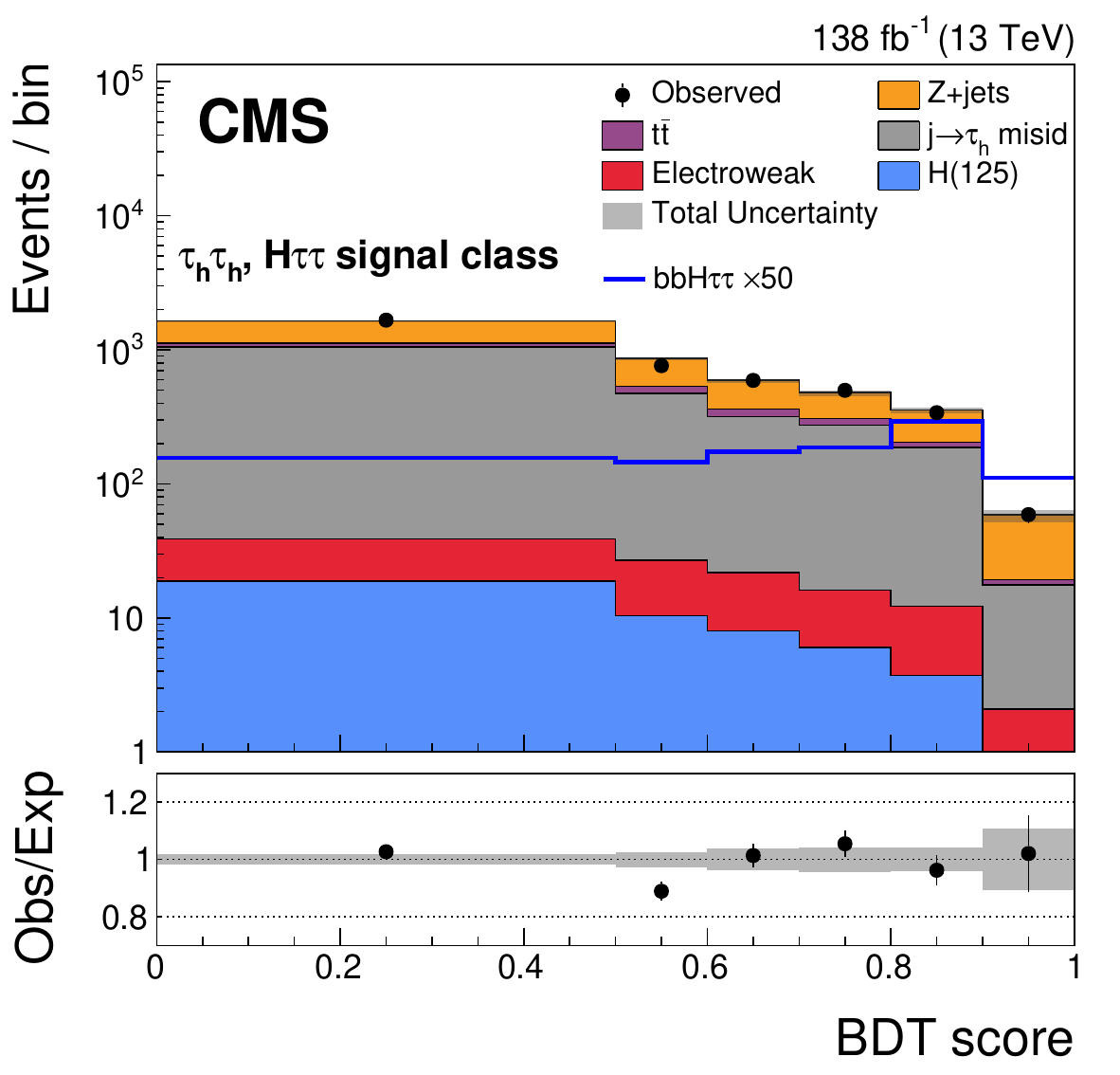}
    \includegraphics[width=.45\textwidth]{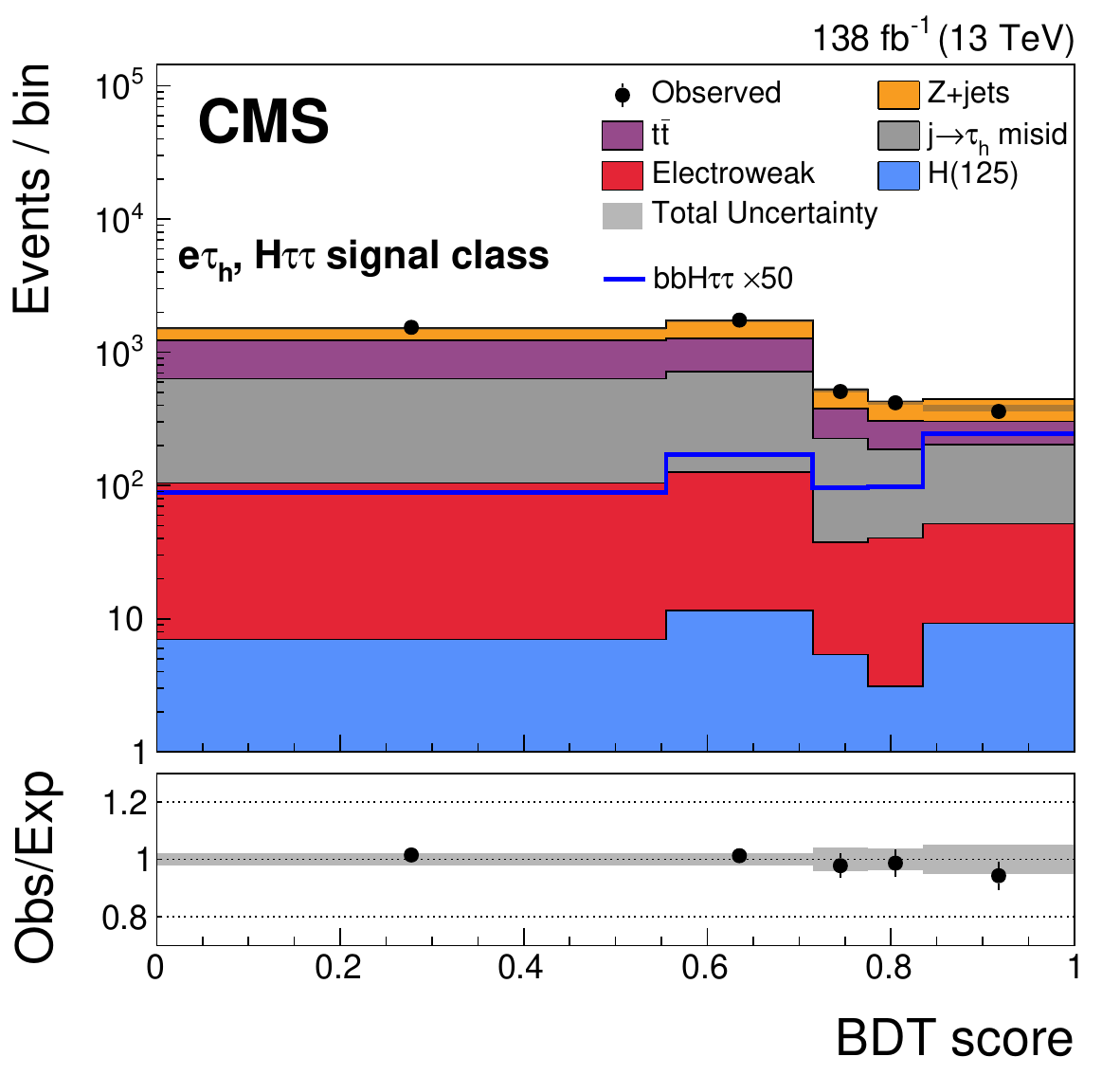}
    \includegraphics[width=.45\textwidth]{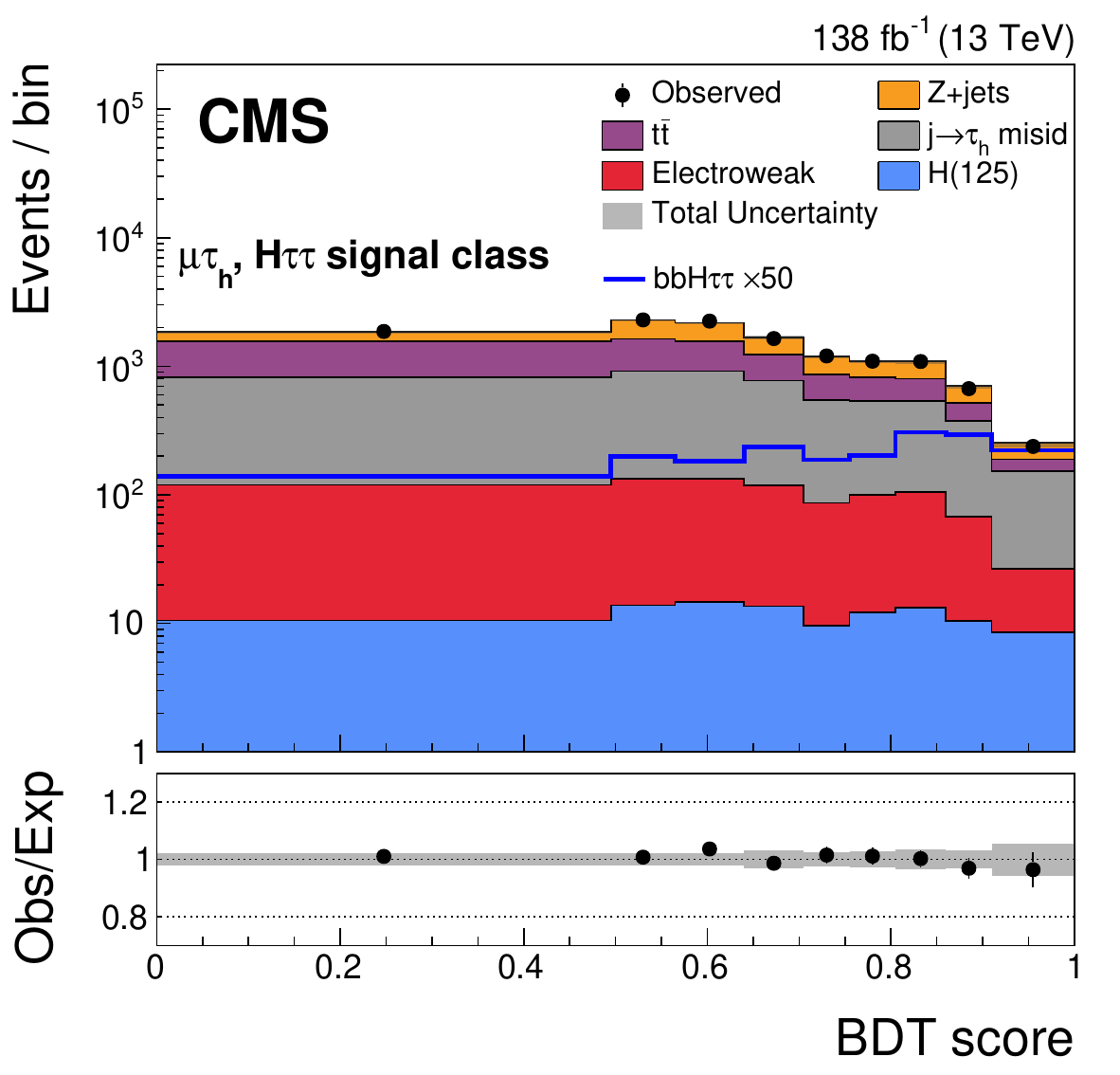}
    \includegraphics[width=.45\textwidth]{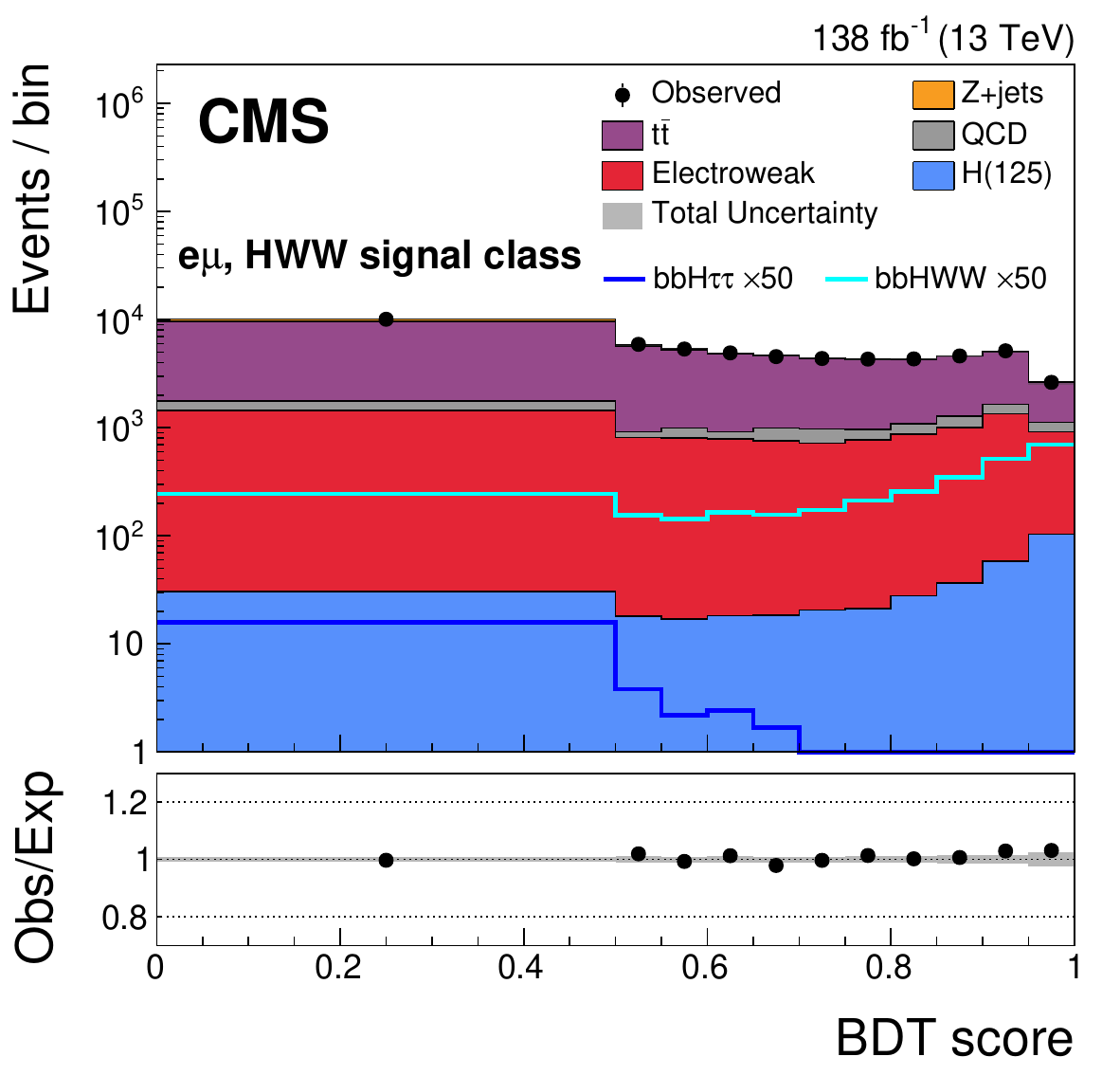}
  \caption{The BDT \HTT class output score distributions for the \tautau (upper left), \etau (upper right), and \mutau (lower left) channels; and the \HWW output score for the \emu channel (lower right). The bin widths in the different channels were optimized independently, considering the analysis sensitivity as well as the size of statistical uncertainties in the simulated templates. The \bbH signal is multiplied by a factor of 50, while all other processes are scaled according to a combined fit of all BDT categories for all channels and years used in this analysis. The total uncertainty includes the statistical and systematic uncertainties. Electroweak processes in the figure include diboson, \Wjets, and single top quark production. For channels involving \tauh candidates, the $\jtotauh$ misid contribution is estimated from data with the \FF method and grouped together. Simulated events with jets misidentified as \tauh candidates are removed from the electroweak, \DYjets, and \ttbar groups. For the \emu channel, the QCD multijet process is estimated using the ``ABCD'' method. The \PH{}(125) group includes processes where a Higgs boson is produced not in association with \PQb quarks, including the top quark associated production and Higgs-strahlung processes, since the \PQb jets in these events originate from the top quark and vector boson decays.}
  \label{fig:BDT_prob}
\end{figure*}

As discussed in the next sections, the limits on the \bbH production cross section are extracted from a combined fit of the distributions of the BDT scores across all categories. 
Figure~\ref{fig:BDT_prob} shows these distributions for the \HTT categories in the \tautau, \etau, and \mutau channels, and for the \HWW category in the \emu channel. The lower-BDT score regions are dominated by background processes, while the higher-score regions show an increasing contribution from the \bbH process.

\section{Systematic uncertainties}
\label{sec:8Syst}
The uncertainty model includes theoretical uncertainties, experimental uncertainties, and uncertainties due to the limited size of the simulated samples of events.

Theoretical uncertainties include those in the parameters used to compute the production cross section predictions. These are uncertainties in the chosen value of the strong coupling, the uncertainty coming from higher-order QCD and EW corrections, and effects due to the variation of the renormalization and factorization scales.
Uncertainties from higher-order corrections to the \bbH production cross section lead to a normalization uncertainty ranging from $-24\%$ to 20\% for the \bbHyb and \bbHybyt processes, and $-6.7\%$ to 4.6\% for \ggH production. An additional uncertainty ranging from $-31\%$ to 47\% is introduced to cover theoretical uncertainties in the additional gluon splitting to \bbar (\bbHyt)~\cite{LHCHiggsCrossSectionWorkingGroup:2016ypw,Deutschmann:2018avk}.

The scale variation uncertainties not only affect the process normalization, but also the shapes of the distributions used for the statistical inference. These variations are computed by independently multiplying and dividing these scales by a factor of two with respect to their theoretical predictions, and re-evaluating the \bbH production cross section. Additional uncertainties are introduced to cover the scale variation effects on the parton shower.

Uncertainties that affect only the normalization of the distributions used for the statistical inference follow a log-normal probability density function, and are labelled `lnN' in Table~\ref{tab:uncertainties}. Some of the most notable are: 
\begin{itemize}
    \item the uncertainty in the integrated luminosity measurement, partially correlated between the different data-taking periods. The integrated luminosities for the 2016, 2017, and 2018 data-taking years have individual uncertainties ranging from 1.2 to 2.5\%~\cite{CMS-LUM-17-003,CMS-PAS-LUM-17-004,CMS-PAS-LUM-18-002}, while the overall uncertainty for the 2016--2018 period is 1.6\%;
    \item the electron and muon identification efficiency uncertainties are estimated to be approximately 2\%;
    \item the uncertainties in the misidentification rate of electrons and muons to \tauh depend on the $\tauh \eta$, and are estimated to be 7.5\% and 6\%, respectively; 
    \item the production cross section uncertainties are estimated as 5\% for the diboson and single top quark processes~\cite{Campbell:2011bn,Gehrmann:2014fva}, 6\% for \ttbar~\cite{Czakon:2011xx,Kidonakis:2013zqa}, 4\% for the \Wjets process~\cite{Melnikov:2006kv}, 2\% for \DYjets~\cite{Melnikov:2006kv,FEWZ}, and 15\% for \ttbar{}\Vjets~\cite{CMS:2017ugv}.
\end{itemize}

\begin{table*}[h!]
    \topcaption{Summary table of the systematic uncertainties affecting the background processes. For uncertainties that vary significantly depending on the kinematic properties of the event, a range indicates the typical size of the uncertainty. The labels `lnN' and `shape' are used, respectively, for uncertainties affecting only the process normalization or having a shape-altering effects.}
    \centering
	\renewcommand{\arraystretch}{1.05}
    \begin{tabular}{L{5cm}  c  c  c}
    Description                   	& Value&  Templates affected &     Type      \\
    \hline
    \multirow{3}{*}{Luminosity}                  	& 2016: 1.2\% &  \multirow{3}{*}{MC} &      \multirow{3}{*}{lnN}     \\
    & 2017: 2.3\%&&\\
    & 2018: 2.5\%&&\\[\cmsTabSkip]
    \multirow{7}{*}{Production cross section} & 2\% & DY & lnN\\
     & 6\% & \ttbar & lnN\\
     & 4\% & \Wjets & lnN\\
     & 5\% & VV & lnN\\
      & 5\% & single top & lnN\\
     & 15\% & \ttbar{}\Vjets & lnN\\
     & 0.5--8\% & H (except \bbH) & lnN\\[\cmsTabSkip]
    \HTT branching fraction & 2.1\% & \HTT  & lnN\\
    \HWW branching fraction & 1.5\% & \HWW & lnN\\[\cmsTabSkip]
    \alpS variation & 3.2\% & \bbH &lnN\\
    \PGm/\Pe identification		& 2\%.  &  MC & lnN           \\
    $\emu$ trigger                       	& 1.5\% & MC &  lnN         \\
    Single $\mu/e$ trigger                       	& \pt and $\eta$ dep. ($<$1\%) & MC &  shape         \\
    \tauh trigger                       	& \pt and DM dep. ($<$5\%) & MC &  shape         \\
    b tagging                    	& 1--9\% &   MC &    shape       \\
    $\PGm(\Pe)\to\tauh$ fake rate             & $\upeta_{\tauh}$ dep. (1--7\%) &   MC with $\ell\to\tauh$ & shape          \\
    \tauh identification                    	   	& \pt and DM dep. (2--3\%)&  MC &     shape      \\
    \tauh energy scale      	& DM dep. (1\%) & MC  &   shape        \\[\cmsTabSkip]
    Jet energy scale              	& 5--10\% &  MC &   shape        \\
    Jet energy resolution             	& 2--5\% &  MC &   shape        \\
    \ptmiss unclustered energy scale & 5--10\% &  MC &   shape        \\[\cmsTabSkip]
    Top quark \pt reweighting       	& $<$10\% & \ttbar, single top   &   shape        \\
    \PZ boson \pt reweighting    & $<$5\%&  DY &   shape        \\[\cmsTabSkip]
    QCD multijet unc.          & 10--15\%  &   nonprompt $\ell$ &  shape         \\
    \FF uncertainties          & 1--5\%  &   $\jtotauh$ fakes &  shape         \\[\cmsTabSkip]
    Prefiring        & 0.5--1\%&  MC &    lnN       \\
    Bin-by-bin stat. unc.        & $\sqrt{{N_{\text{events}}}}$ &   All &  shape         \\
    \end{tabular}
    \label{tab:uncertainties}
\end{table*}

Uncertainties in the energy scale of objects such as jets, \tauh, and \ptmiss alter the distributions of the input features to the BDT models, and as such modify the shape of the BDT score distribution. The jet energy is corrected based on sources such as the detector efficiency during data taking, pileup interactions, and the clustering efficiency. The corresponding uncertainties are divided into groups depending on whether they are correlated across data-taking years and channels, or uncorrelated. They are shown in Table~\ref{tab:uncertainties} as \textit{Jet ES} and \textit{Jet energy resolution}, and their values depend primarily on the jet \pt. These corrections are propagated to the \ptmiss, in addition to those accounting for PF candidates that are not clustered within jets.
The \tauh identification and energy scale depend on the \tauh candidate decay mode (DM) and \pt. Additional uncertainties cover the energy scales of prompt and misidentified electrons and muons, and the \ptmiss.

The estimation of background contributions originating from nonprompt or misidentified leptons and jets leads to additional uncertainties. 
The ``ABCD'' method for the \emu channel and the \FF method for the \tautau, \etau, and \mutau channels introduce shape-altering uncertainties pertaining to the statistical uncertainties in the samples used for the measurement of the extrapolation scale factors, the misidentification rate for jets as leptons or \tauh, and closure tests on observables used in the analysis. Nonclosure uncertainties, generally below 5--7\% across the studied parameter space, cover the residual disagreement in the background modelling.

The shape correction for the \PQb tagging classifier introduces shape-altering effects on the BDT score. Further uncertainties originate from the correction of the top quark and \PZ boson \pt spectra in the \ttbar~\cite{arXiv:2402.08486} and \DYjets processes, respectively. For both, the uncertainty is estimated as the size of the correction itself. 

During the 2016 and 2017 data-taking periods, an increase in the offset of the ECAL timing pulse led the Level-1 trigger relying on ECAL to fire on the previous bunch crossing (prefiring). An uncertainty, negligible in some cases and ranging up to 4\%, is introduced to account for this phenomenon.

Uncertainties accounting for the finite sizes of the simulated event samples, and the sideband regions used for the data-driven background estimates, are incorporated via the Barlow--Beeston ``light'' procedure~\cite{Barlow:1993dm,Conway:2011in}.

\section{Results}
\label{sec:9Results}

The statistical analysis is performed via a simultaneous binned maximum likelihood fit to the data in all BDT categories described in Section~\ref{sec:7BDT}. The results were obtained using the CMS statistical analysis tool \textsc{Combine}~\cite{CAT-23-001}.

Upper limits at the 95\% confidence level (\CL) are set, considering as signal the sum of the \bbHyb and \bbHyt contributions together with their interference term. These sum up to a total cross section of $\sigma_{\text{theory}}= 1.489\unit{pb}$. Limits are placed on the ratio between the measured cross section and its theoretical prediction:
\begin{equation}
    \mu=\frac{\sigma(\Pp\Pp\to \bbH(y_\PQb,y_\PQt))}{\sigma_{\text{theory}}},
\end{equation}
which is referred to as the signal strength.

The upper limits are set using the modified frequentist \CLs criterion~\cite{CLS2,Read:2002hq}, with the profile likelihood ratio modified for upper limits~\cite{CMS-NOTE-2011-005} as the test statistic. The asymptotic approximation~\cite{CowanLikelihood} is used in the limit setting procedure. To extract the two-dimensional constraints presented later in this section, we use the profile likelihood ratio from Ref.~\cite{CMS-NOTE-2011-005} as the test statistic. The 68\% (95\%) confidence intervals are constructed as the union of points for which the difference in twice the negative log-likelihood with respect to the minimum is below 2.28 (5.99). 
In all cases, the experimental and theoretical uncertainties are incorporated in the likelihood as nuisance parameters.

\begin{figure*}[h!]
    \centering
    \includegraphics[width=.65\textwidth]{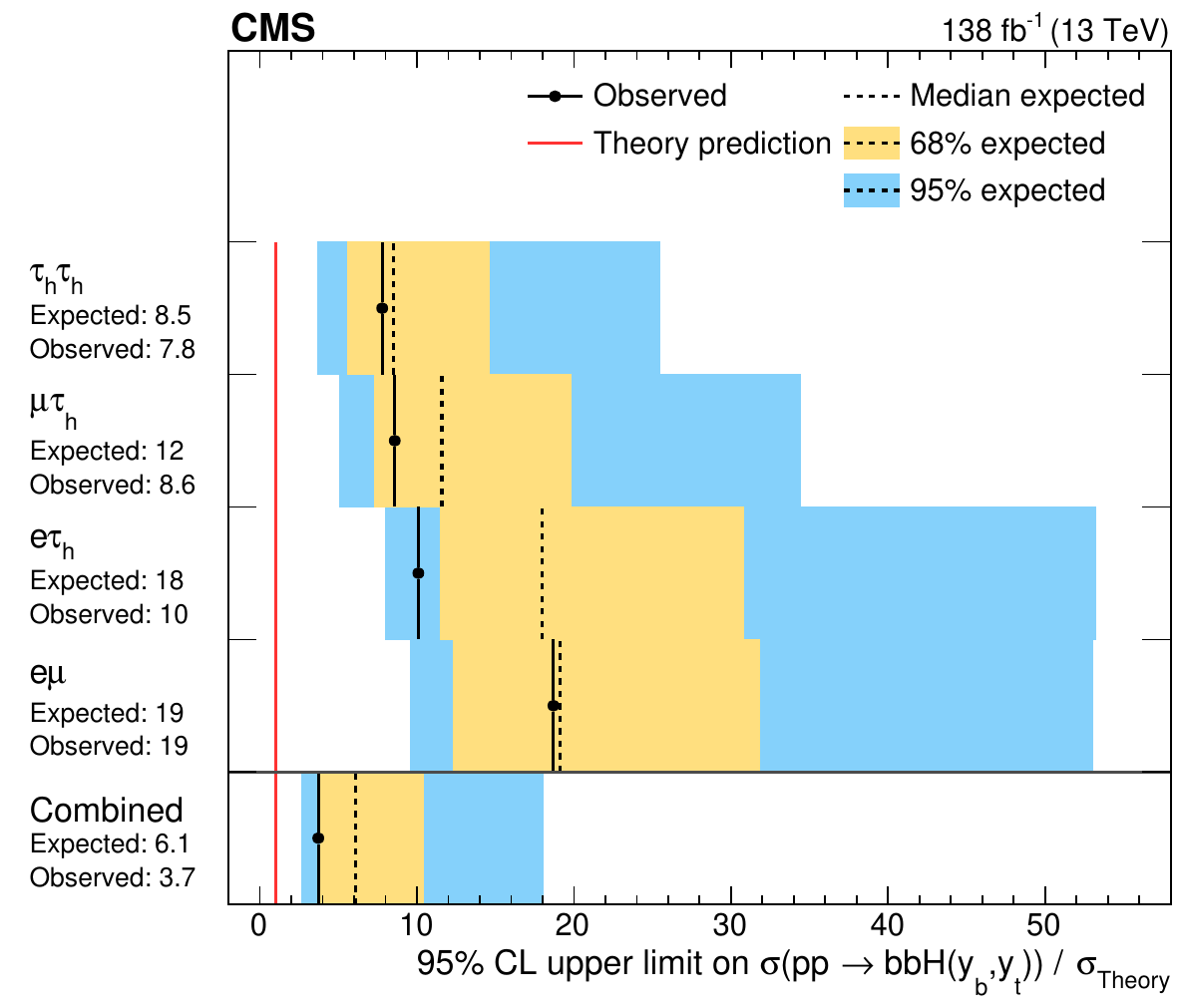}
    \caption{Upper limits at the 95\% \CL on the signal strength for the the $\Pp\Pp\to \bbH(y_\PQb,y_\PQt)$ process. The terms in which the Higgs boson is produced via Yukawa couplings with top or bottom quarks contribute to the estimated relative production cross sections. The interference term between these contributions is also accounted for. The $\Pp\Pp\to \PZ(\to \bbar)\PH$ process is treated as a background in this search. The theoretical prediction, shown as a red line placed at 1, corresponds to the estimated production cross section of 1.489\unit{pb}. The black markers show the observed limits, and the dashed lines with the yellow and blue uncertainty bands represent the expected upper limits with their 68\% and 95\% central intervals.}
    \label{fig:UpperLimits}
\end{figure*}

The 95\% \CL upper limits on the signal strength are shown in Fig.~\ref{fig:UpperLimits}. 
These limits are obtained by coherently varying all contributions to the production cross section considered, namely those depending on Yukawa couplings to the bottom or top quark and their interference term. The combination of all channels yields an observed (expected) upper limit of 3.7 (6.1) at 95\% \CL.

Varying each contribution separately makes it possible to constrain the coupling structure of the Higgs boson. This is done by introducing the coupling scaling parameters \kapt and \kapb, which represent deviations from the SM expectation in the Higgs boson coupling strengths, and performing a likelihood ratio scan over the \kapt-\kapb parameter space.
To account for the bottom quark contribution to the quark loop in the \bbHyt process, this contribution is scaled by $1.04\kapt^2-0.04\kapb\kapt+0.002\kapb^2$~\cite{LHCHiggsCrossSectionWorkingGroup:2016ypw}, while the \bbHyb contribution and the interference term are scaled by $\kapb^2$ and $\kapb\kapt$, respectively. 
To constrain the \kapt parameter, the results obtained in this analysis have been combined with the Higgs boson production cross section measurement in final states with two tau leptons~\cite{CMS:2022kdi} previously published by the CMS Collaboration. The published measurement required the absence of \PQb-tagged jets, meaning that its limits on \kapb are entirely derived from the constraints on the Higgs boson decay width that are present in the coupling modifier model. The combination was performed at the level of the inputs for the statistical inference, because the presence of a \PQb jet veto in Ref.~\cite{CMS:2022kdi} makes that measurement orthogonal to the search presented here.

\begin{figure*}[!htb]
  \centering
    \includegraphics[width=.85\textwidth]{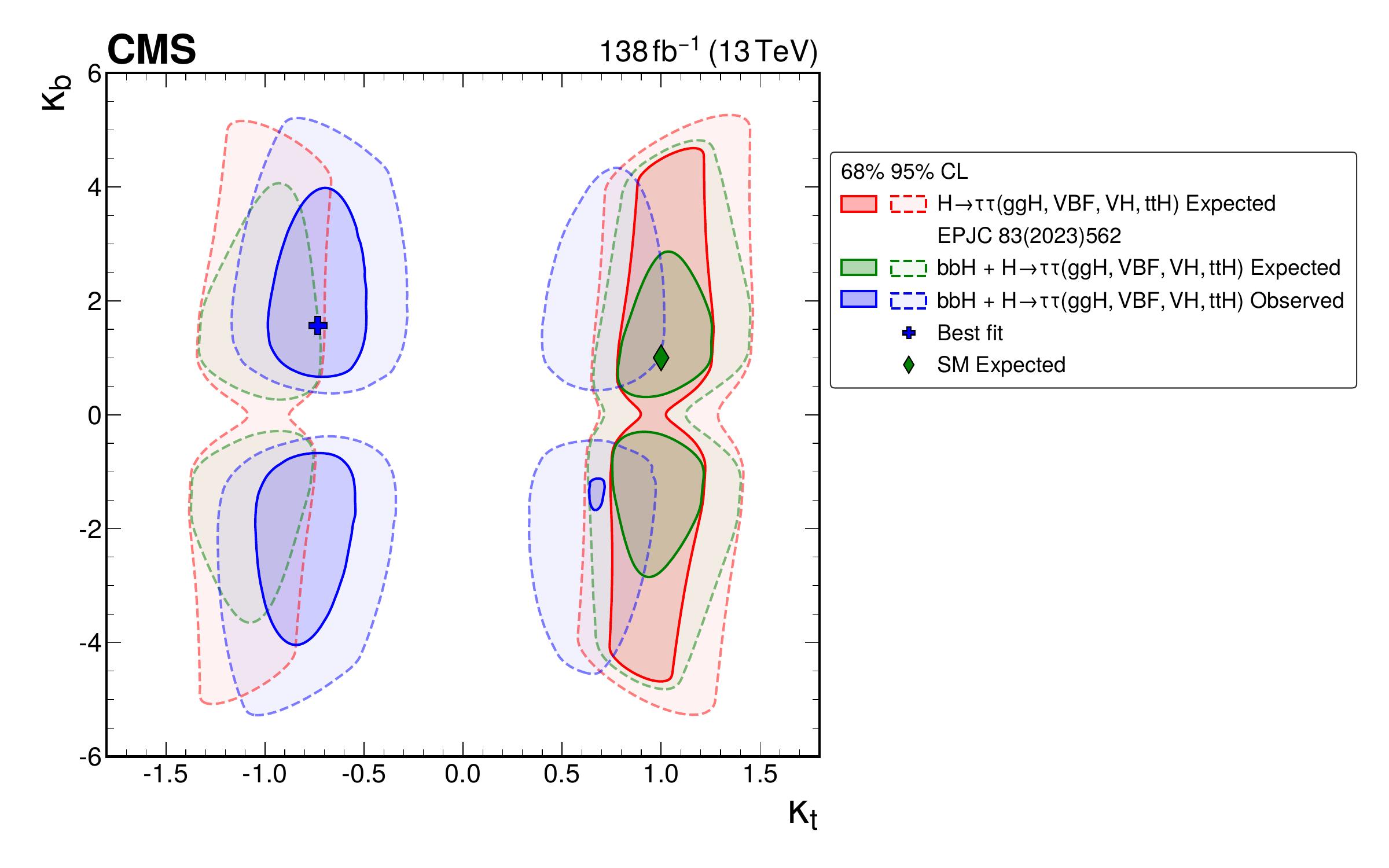}
  \caption{Two-dimensional confidence intervals on the $\kapb$ and $\kapt$ parameters for the channels studied in this search. Expected limits are shown in red for the \HTT cross section measurement \cite{CMS:2022kdi} performed for other Higgs boson production mechanisms and in green for the combination with the  analysis presented in this Letter. The observed constraints are shown in blue, with a cross marking the best fit point. A green diamond is placed to mark the SM expectation. Solid lines with shaded areas mark the 68\% confidence interval contours, and dashed lines mark the 95\% confidence interval.}
  \label{fig:kappa_limit_tau}
\end{figure*}

The expected constraints in the \kapt-\kapb parameter space are shown in Fig.~\ref{fig:kappa_limit_tau} for the previous CMS results in red, and in combination with the study presented in this Letter in green. A noticeable improvement in the constraint on the \kapb parameter, shown by the tightening of the 68\% confidence interval contour, is observed in the combined limits.
The constraints observed when fitting to the data are shown in blue, with the best fit point found at the coordinates $(\kapt,\kapb)=(-0.73,1.58)$. The limits are compatible with the SM expectation at 95\% \CL.
In this fit the \kapb and \kapt parameters are left freely floating, together with the coupling modifier for the tau leptons Yukawa coupling ($\kappa_\PGt$); the other coupling parameters are fixed to their SM values. This means some of the constraints on \kapb are indirect, from assumptions on the total Higgs boson decay width. The effects of these constraints can be observed in the red contours in Fig.~\ref{fig:kappa_limit_tau}, as the measurement of the \HTT production cross section from Ref.~\cite{CMS:2022kdi} did not include Higgs boson production processes involving the \kapb coupling, beyond its contribution to the quark loop in the \ggH process.

Most of the considered processes are not sensitive to the sign of the Yukawa couplings, leading to a partial degeneracy with respect to the coupling sign. This is more noticeable for \kapb, since the degeneracy is only broken by the \bbHybyt interference term and the different flavour contributions to the quark loop in the \ggH process.  
The observed limits on \kapt show a slight preference for negative values and are compatible with the SM prediction at 95\% \CL.
The observed limits on \kapb exclude $\kapb=0$ at 95\% \CL and are also compatible within 68\% \CL with those measured in a combined fit of Higgs boson production and decay channels, including $\PH\to\bbar$, performed by the CMS Collaboration with data collected in Run 2~\cite{CMS:Hig102022}. At present, the uncertainty in the \kapb measurement is around 7 times larger than the established constraints provided by the analysis of Higgs boson decays.

\section{Summary}
\label{sec:10Summary}
A search for the 125\GeV Higgs boson produced in association with bottom quarks and decaying into a pair of tau leptons or \PW bosons has been presented. The search was performed on data collected by the CMS experiment in the period 2016--2018 at a centre-of-mass energy of $\sqrt{s}=13\TeV$, corresponding to an integrated luminosity of 138\fbinv. This search was performed in four final states: \tautau, \etau, \mutau, and \emu. Higgs boson decays to tau leptons were targeted in all four final states, while $\PH\to\PW\PW$ decays contributed only in the \emu channel as a result of the kinematical similarities between the two decay processes. At the current level of precision, the background processes provide an adequate description of the observed data, and no significant excess above the background-only expectation was found. The observed (expected) upper limit at the 95\% confidence level (\CL) on the joint \bH and \bbH production cross section is 3.7 (6.1) times the standard model prediction. The search also constrained the Higgs Yukawa couplings to bottom and top quarks in the $\kappa$-model interpretation. The best fit value for the coupling modifiers was found to be $(\kapt,\kapb)=(-0.73,1.58)$. The observed constraints are compatible with the standard model expectation at the 95\% \CL.

\begin{acknowledgments}
  We congratulate our colleagues in the CERN accelerator departments for the excellent performance of the LHC and thank the technical and administrative staffs at CERN and at other CMS institutes for their contributions to the success of the CMS effort. In addition, we gratefully acknowledge the computing centres and personnel of the Worldwide LHC Computing Grid and other centres for delivering so effectively the computing infrastructure essential to our analyses. Finally, we acknowledge the enduring support for the construction and operation of the LHC, the CMS detector, and the supporting computing infrastructure provided by the following funding agencies: SC (Armenia), BMBWF and FWF (Austria); FNRS and FWO (Belgium); CNPq, CAPES, FAPERJ, FAPERGS, and FAPESP (Brazil); MES and BNSF (Bulgaria); CERN; CAS, MoST, and NSFC (China); MINCIENCIAS (Colombia); MSES and CSF (Croatia); RIF (Cyprus); SENESCYT (Ecuador); ERC PRG, RVTT3 and MoER TK202 (Estonia); Academy of Finland, MEC, and HIP (Finland); CEA and CNRS/IN2P3 (France); SRNSF (Georgia); BMBF, DFG, and HGF (Germany); GSRI (Greece); NKFIH (Hungary); DAE and DST (India); IPM (Iran); SFI (Ireland); INFN (Italy); MSIP and NRF (Republic of Korea); MES (Latvia); LMTLT (Lithuania); MOE and UM (Malaysia); BUAP, CINVESTAV, CONACYT, LNS, SEP, and UASLP-FAI (Mexico); MOS (Montenegro); MBIE (New Zealand); PAEC (Pakistan); MES and NSC (Poland); FCT (Portugal); MESTD (Serbia); MCIN/AEI and PCTI (Spain); MOSTR (Sri Lanka); Swiss Funding Agencies (Switzerland); MST (Taipei); MHESI and NSTDA (Thailand); TUBITAK and TENMAK (Turkey); NASU (Ukraine); STFC (United Kingdom); DOE and NSF (USA).
  
  \hyphenation{Rachada-pisek} Individuals have received support from the Marie-Curie programme and the European Research Council and Horizon 2020 Grant, contract Nos.\ 675440, 724704, 752730, 758316, 765710, 824093, 101115353, 101002207, and COST Action CA16108 (European Union); the Leventis Foundation; the Alfred P.\ Sloan Foundation; the Alexander von Humboldt Foundation; the Science Committee, project no. 22rl-037 (Armenia); the Belgian Federal Science Policy Office; the Fonds pour la Formation \`a la Recherche dans l'Industrie et dans l'Agriculture (FRIA-Belgium); the F.R.S.-FNRS and FWO (Belgium) under the ``Excellence of Science -- EOS" -- be.h project n.\ 30820817; the Beijing Municipal Science \& Technology Commission, No. Z191100007219010 and Fundamental Research Funds for the Central Universities (China); the Ministry of Education, Youth and Sports (MEYS) of the Czech Republic; the Shota Rustaveli National Science Foundation, grant FR-22-985 (Georgia); the Deutsche Forschungsgemeinschaft (DFG), among others, under Germany's Excellence Strategy -- EXC 2121 ``Quantum Universe" -- 390833306, and under project number 400140256 - GRK2497, and DASHH (Data Science in Hamburg --- Helmholtz Graduate School for the Structure of Matter) grant HIDSS-0002 (Germany); the Hellenic Foundation for Research and Innovation (HFRI), Project Number 2288 (Greece); the Hungarian Academy of Sciences, the New National Excellence Program - \'UNKP, the NKFIH research grants K 131991, K 133046, K 138136, K 143460, K 143477, K 146913, K 146914, K 147048, 2020-2.2.1-ED-2021-00181, and TKP2021-NKTA-64 (Hungary); the Council of Science and Industrial Research, India; ICSC -- National Research Centre for High Performance Computing, Big Data and Quantum Computing and FAIR -- Future Artificial Intelligence Research, funded by the NextGenerationEU program (Italy); the Latvian Council of Science; the Ministry of Education and Science, project no. 2022/WK/14, and the National Science Center, contracts Opus 2021/41/B/ST2/01369 and 2021/43/B/ST2/01552 (Poland); the Funda\c{c}\~ao para a Ci\^encia e a Tecnologia, grant CEECIND/01334/2018 (Portugal); the National Priorities Research Program by Qatar National Research Fund; MCIN/AEI/10.13039/501100011033, ERDF ``a way of making Europe", and the Programa Estatal de Fomento de la Investigaci{\'o}n Cient{\'i}fica y T{\'e}cnica de Excelencia Mar\'{\i}a de Maeztu, grant MDM-2017-0765 and Programa Severo Ochoa del Principado de Asturias (Spain); the Chulalongkorn Academic into Its 2nd Century Project Advancement Project, and the National Science, Research and Innovation Fund via the Program Management Unit for Human Resources \& Institutional Development, Research and Innovation, grant B39G670016 (Thailand); the Kavli Foundation; the Nvidia Corporation; the SuperMicro Corporation; the Welch Foundation, contract C-1845; and the Weston Havens Foundation (USA).  
\end{acknowledgments}
\bibliography{auto_generated} 
\cleardoublepage \appendix\section{The CMS Collaboration \label{app:collab}}\begin{sloppypar}\hyphenpenalty=5000\widowpenalty=500\clubpenalty=5000
\cmsinstitute{Yerevan Physics Institute, Yerevan, Armenia}
{\tolerance=6000
A.~Hayrapetyan, A.~Tumasyan\cmsAuthorMark{1}\cmsorcid{0009-0000-0684-6742}
\par}
\cmsinstitute{Institut f\"{u}r Hochenergiephysik, Vienna, Austria}
{\tolerance=6000
W.~Adam\cmsorcid{0000-0001-9099-4341}, J.W.~Andrejkovic, T.~Bergauer\cmsorcid{0000-0002-5786-0293}, S.~Chatterjee\cmsorcid{0000-0003-2660-0349}, K.~Damanakis\cmsorcid{0000-0001-5389-2872}, M.~Dragicevic\cmsorcid{0000-0003-1967-6783}, P.S.~Hussain\cmsorcid{0000-0002-4825-5278}, M.~Jeitler\cmsAuthorMark{2}\cmsorcid{0000-0002-5141-9560}, N.~Krammer\cmsorcid{0000-0002-0548-0985}, A.~Li\cmsorcid{0000-0002-4547-116X}, D.~Liko\cmsorcid{0000-0002-3380-473X}, I.~Mikulec\cmsorcid{0000-0003-0385-2746}, J.~Schieck\cmsAuthorMark{2}\cmsorcid{0000-0002-1058-8093}, R.~Sch\"{o}fbeck\cmsorcid{0000-0002-2332-8784}, D.~Schwarz\cmsorcid{0000-0002-3821-7331}, M.~Sonawane\cmsorcid{0000-0003-0510-7010}, W.~Waltenberger\cmsorcid{0000-0002-6215-7228}, C.-E.~Wulz\cmsAuthorMark{2}\cmsorcid{0000-0001-9226-5812}
% ...
\par}
\vskip\cmsinstskip
\dag:~Deceased\\
$^{1}$Also at Yerevan State University, Yerevan, Armenia\\
$^{2}$Also at TU Wien, Vienna, Austria\\
% ...
\end{sloppypar}
%%% END EDITABLE REGION %%%
% skeleton_end
\end{document}